\documentstyle[array,aps]{revtex}
\begin{document}
\draft
\title
{Entanglement measure for any quantum states}
\author{Hyuk-jae Lee$^1${\footnote{e-mail:lhjae@iquips.uos.ac.kr}},
Sung Dahm Oh$^2${\footnote{e-mail:sdoh@sookmyung.ac.kr}} and
Doyeol Ahn$^{1,3}${\footnote{e-mail:dahn@uoscc.uos.ac.kr}}}
\address{
$^1$Institute of Quantum Information Processing and Systems,
University of Seoul, Seoul, 130-743, Korea\\
$^2$ Department of Physics, Sookmyung Women's University, Seoul,140-742, Korea\\
$^3$Department of Electrical and computer Engineering,
University of Seoul,
Seoul, 130-743, Korea\\
}

%
\maketitle


\baselineskip24pt



\begin{abstract}
The entanglement measure for multiqudits is proposed. This measure
calculates the partial entanglement distributed by subsystems and
the complete entanglement of the total system. This shows that we
need to measure the subsystem entanglements to explain the full
description for multiqudit entanglement. Furthermore, we extend
the entanglement measure to mixed multiqubits and the higher
dimension Hilbert spaces.
\end{abstract}
\vspace{.25in}


\newpage

Entanglement has been recently recognized as the most essential
ingredient in the quantum information technology as can be seen by
the cases of the superdense coding\cite{benn}, quantum
computation, teleportation\cite{bras}, clock
synchronization\cite{chua}, \cite{abra} and quantum cryptography
\cite{eker}.

In order to clarify the entanglement characteristics, the
entanglement degree has to be quantifiable. The existence of
entanglement for two-qubit cases can be established by negative
partial transpose property of the density matrices
\cite{pere}\cite{horo} and entanglement measures have been
quantized by concurrence\cite{woot}, negativity\cite{vida} and
entanglement of formation\cite{benne}. The mathematical and
physical structures of entanglement have not been yet fully
understood for multipartite cases even in the two dimensional
space. Entanglement in the case of the multipartite systems and
the higher dimensional systems has remained unsolved even though
there were several investigations to quantify the entanglement and
classify types of the entangled states. There have been
investigations on the multipartite entanglement measure by the
{\it tangle}\cite{coff} which computes the concurrence between two
intentionally divided subsystems in effectively two dimensional
Hilbert space. However, tangles cannot properly express the
entanglement degree for any $W$-states. There have been also
recent proposals of an entanglement measure by the operator
norm\cite{yuka} and the hyperdeterminant\cite{mika} of the given
quantum states. However, these schemes cannot explain the full
entanglement structure of the composite system. Here, we propose a
direct measure of the entanglement distributed in the subsystems
and the total system for the given entangled states.

The entanglements in multipartite qubits are complex because the
quantum states can have various types of entanglement sharing
among the subsystems. In order to illuminate this situation,
consider a state of the form
$|\Psi_5\rangle=|Bell\rangle\otimes|GHZ_3\rangle$ in a five-qubit
system, where
$|Bell\rangle=\frac{1}{\sqrt{2}}(|01\rangle-|10\rangle)$, between
the first and the second qubits and $|GHZ_3\rangle$ the
Greenberger-Horne-Zeilinger(GHZ) state among the third, the fourth
and the fifth qubits.  Next, consider a state of the form
$|\Phi_5\rangle=|GHZ_3\rangle\otimes|Bell\rangle$.  Then, both
quantum states have no complete entanglement for the total system
but have different degrees in the specific subsystems.
$|\Phi_5\rangle$ exhibits $GHZ$ entangled in the first, the second
and the third qubits but $|\Psi_5\rangle$ does not show this
property. Until now, there were no any entanglement measures that
can distinguish this situation. Then, entanglement measure for
multipartite systems has to predict the magnitude of all types of
entanglements which exist among the constituents. In this letter,
we present a general entanglement measure for multipartite qubits,
and introduce examples for pure multiqubit systems. Finally, we
explain that our entanglement measure can be extended to
mulitiqubit mixed states and the higher dimensional Hilbert space.

Let us start with the pure states for multipartite qubit system as
\begin{equation}
\Psi(1, 2, 3,\cdots,n)=\sum^1_{ij\cdots k=0}a_{ij\cdots k}
|i\rangle_1 \otimes|j\rangle_2\otimes\cdots\otimes|k\rangle_n,
\label{abc}
\end{equation}
where $n$ denotes the number of qubits and $\sum^1_{ij\cdots
k=0}|a_{ij\cdots k}|^2 =1$. Our question is how to distinguish
separated states from entangled states given in eq. (\ref{abc}).
The three-qubit state, $|\Psi(1,2,3)\rangle$, has two types of
entanglement different from two qubits system. The first is the
entanglement between two particles and the second is among three
particles. Furthermore, entanglement between two particles has
three possibilities, depending on which qubit is separated. The
increase of the qubit numbers in the system produces the increase
of the possibilities in the entanglement types. Then, we have to
differentiate all these situations if we suggest the entanglement
measure.

For this purpose, we consider the quantum correlations for the
given multipartite qubit system $|\Psi\rangle$ as the following;
\begin{equation}
M_{ijkl\cdots}(\alpha, \beta,
\gamma,\cdots;|\Psi\rangle)=\langle(\sigma_i(\alpha)-\lambda_i(\alpha))\otimes
(\sigma_j(\beta)-\lambda_j(\beta))\otimes(\sigma_k(\gamma)-\lambda_k(\gamma))\otimes\cdots\rangle,
\label{mtensor}
\end{equation}
where $\sigma_i(\alpha)$ denotes the $i^{th}$-component Pauli's
matrix of $\alpha$-th qubit and $\lambda_i(\alpha)=\langle
I(1)\otimes I(2)\otimes I(3)\cdots \sigma_i(\alpha)\otimes
I((\alpha+1)\otimes\cdots\rangle$. Here, $I(\alpha)$ is the identity
operator in $\alpha$-th qubit. We will show that $M$ is zero for
the completely separable state later. Define the tensor form as
\begin{eqnarray}
M'_{ijkl\cdots}(\alpha, \beta,
&&\gamma,\cdots;|\Psi\rangle)= M_{ijkl\cdots}(\alpha, \beta, \gamma,\cdots;|\Psi\rangle)\nonumber\\
&&-\sum (\mbox{all the possible partitions of indices of } M_{ijkl\cdots}(\alpha,
\beta, \gamma,\cdots;|\Psi\rangle)).
\label{mtensorp}
\end{eqnarray}
The sum of the second term in the right side of eq.
(\ref{mtensorp}) appears in the case of the system which is
composed by more than three qubits. In the four-qubit case,
\begin{eqnarray}
M'_{ijkl}(1,2,3,4;|\Psi\rangle)&=&M_{ijkl}(1,2,3,4;|\Psi\rangle)-M_{ij}(1,2;|\Psi\rangle)M_{kl}(3,4;|\Psi\rangle)\nonumber\\
&&-M_{ik}(1,3;|\Psi\rangle)M_{jl}(2,4;|\Psi\rangle)-M_{il}(1,4;|\Psi\rangle)M_{jk}(2,3;|\Psi\rangle).
\end{eqnarray}

Define the entanglement measure from $M'$;
\begin{equation}
B^{(m)}(\alpha, \beta,\cdots \gamma;|\Psi\rangle)=\frac{1}{{\cal N}}\sum_{ijkl\cdots}M'_{ijkl\cdots}(\alpha, \beta,
\cdots\gamma;|\Psi\rangle)M'_{ijkl\cdots}(\alpha, \beta,
\cdots\gamma;|\Psi\rangle),
\label{entmeasure}
\end{equation}
where ${\cal N}$ is a normalization constant which depends on the
number of qubits, $m$. $B^{(m)}(\alpha, \beta,
\cdots\gamma;|\Psi\rangle)$ calculates the entanglement magnitude
among $m$ qubits labelled by $\alpha, \beta, \cdots\gamma$. For
example, $B^{(2)}(\alpha, \beta;|\Psi\rangle)$ describes the
entanglement magnitude between $\alpha$ and $\beta$ qubits and
$B^{(3)}(\alpha, \beta, \gamma;|\Psi\rangle)$ the entanglement
degree among $\alpha$, $\beta$ and $\gamma$ qubits and so on. In
two-qubit systems, there only exists $B^{(2)}(1,2;|\Psi\rangle)$
which is the same measure as Schlinz and Mahler's entanglement
measure\cite{mahl}. Measures of eq. (\ref{entmeasure}) satisfy the
following properties:
\begin{itemize}
\item $B^{(m)}=0$ for completely separable states.
\item $B^{(m)}\geq 0$.
\item $B^{(m)}$ is invariant under any local unitary transformations.
\end{itemize}
The first property can be shown easily with simple calculations
since $\langle(\sigma_i(\alpha)-\lambda_i(\alpha))\otimes
(\sigma_j(\beta)-\lambda_j(\beta))\otimes(\sigma_k(\gamma)-\lambda_k(\gamma))\otimes\cdots\rangle=(\langle
\sigma_i(\alpha)\rangle-\lambda_i(\alpha))
(\langle\sigma_j(\beta)\rangle-\lambda_j(\beta))(\langle\sigma_k(\gamma)\rangle-\lambda_k(\gamma))
\cdots$, in completely separate states. The second property is
true since $M=0$ in completely separable states and $B$ is defined
by the square of real numbers. The third property is shown easily
by using $U^{\dagger}\sigma_i U=T_{ij}\sigma_j$ and
$\sum_{i}T_{ij}T_{ik}=\delta_{jk}$ where $U$ is an unitary matrix
and $T$ is a $3\times 3$ orthogonal matrix\cite{mahl}.

Let us explain entanglement degrees for two-, three- and
four-qubit cases through direct calculations. For pure two-qubit,
an arbitrary pure entangled state can be written by
$|\Psi_2\rangle= a |00\rangle + b |11\rangle$ where $a$ and $b$
are the nonnegative real number coefficients with normalization
$|a|^2+|b|^2=1$ appearing in the Schmidt's decomposition. The
entanglement of this state can be calculated simply by using eq.
(\ref{entmeasure}). Then, $B^{(2)}(1,2;|\Psi_2\rangle)=4a^2
b^2=C^2$ where $C$ is the concurrence. This explains that
$B^{(2)}(1,2;|\Psi_2\rangle)$ is a monotonically increasing
function of $C$ in the region of $0\leq C\leq 1$. Then, one can
see that our measure is appropriate in bipartite qubits.

In the three-qubit cases, we consider the two complete
entanglements such as $|GHZ_3\rangle$ and $|W_3\rangle$ states and
a partially entangled state as $|\Psi_3\rangle=|Bell\rangle
\otimes |0\rangle$, that the first and the second qubits are
entangled and the third qubit is separated. We are summarizing the
obtained results for all the possible three-qubit states in the
following table:

\vspace{0.7cm}
\begin{tabular}{|c|c|c|}
\hline\hline
{}&$B^{(2)}(\alpha, \beta;|\Psi\rangle)$ & $B^{(3)}(1,2,3;|\Psi\rangle)$\\
\hline
$|GHZ_3\rangle$ & $\frac{1}{3}$ & 1\\
\hline
$|W_3\rangle$ & $\frac{88}{243}$ & $\frac{280}{729}$\\
\hline
$|\Psi_3\rangle$ & $1$ or $0$ & 0\\
\hline\hline
\end{tabular}
\vspace{0.7cm}

Here, $1$ or $0$ in $B^{(2)}$ of $|\Psi_3\rangle$, represents
$B^{(2)}(1,2;|\Psi_3\rangle)=1$, and
$B^{(2)}(1,3;|\Psi_3\rangle)=B^{(2)}(2,3;|\Psi_3\rangle)=0$,
respectively. This explains the situation of entanglement for
individual subsystems in $|\Psi_3\rangle$ well. The entanglement
of $|W_3\rangle$ is stronger than $|GHZ_3\rangle$ between two
particles but the entanglement of $|W_3\rangle$ is weaker than
$|GHZ_3\rangle$ for the complete entanglement. Note that the
bipartite entanglement does not disappear in GHZ state. It has
been generally accepted that the partial entanglement degree of
$(N-1)$ qubits of $|GHZ_N\rangle$ is zero with an artificial
reduction of a qubit. We here point out that it is inappropriate
to view the partial entanglement, tracing out on qubit.

For four-qubit, let us treat four four-qubit quantum states such
as $|GHZ_4\rangle$, $|W_4\rangle$,
$|\phi_6\rangle=\frac{1}{\sqrt{6}}(|0011\rangle+|0101\rangle+|1001\rangle+|1010\rangle+|0110\rangle+|1100\rangle)$,
and
$|\phi_4\rangle=\frac{1}{2}(|0000\rangle+|0011\rangle+|1100\rangle-|1111\rangle)$
which are completely entangled. We also calculate $B^{(m)}$ for a
partially entangled states, $|GHZ_3\rangle\otimes |0\rangle$ and
$|Bell\rangle\otimes|Bell\rangle$.

\vspace{0.7cm}
\begin{tabular}{|c|c|c|c|}
\hline\hline
{}&$B^{(2)}(\alpha, \beta;|\Psi\rangle)$ & $B^{(3)}(\alpha,\beta,\gamma;|\Psi\rangle)$&$B^{(4)}(1,2,3,4;|\Psi\rangle)$\\
\hline
$|GHZ_4\rangle$ & $\frac{1}{3}$ & $0$ & 1\\
\hline
$|W_4\rangle$ & $\frac{3}{16}$ & $\frac{7}{64}$ & $\frac{51}{256}$\\
\hline
$|\phi_6\rangle$ & $\frac{1}{3}$   & $0$ & $\frac{7}{27}$\\
\hline
$|\phi_4\rangle$ & $\frac{1}{3}$ or $0$ & $0$&$\frac{1}{3}$\\
\hline
$|GHZ_3\rangle\otimes |0\rangle$ & $\frac{1}{3}$ & $1$ or $0$ & 0\\
\hline
$|Bell\rangle\otimes|Bell\rangle$ & $1$ or $0$ & $0$ & $0$\\
\hline\hline
\end{tabular}
\vspace{0.7cm}

$\frac{1}{3}$ or $0$ in $B^{(2)}$ of $|\phi_4\rangle$ means that
$B^{(2)}(1, 2;|\phi_4\rangle)=B^{(2)}(3,
4;|\phi_4\rangle)=\frac{1}{3}$ and $B^{(2)}$ for any other
combinations of two qubits are zero, and $1$ or $0$ in $B^{(2)}$
of $|Bell\rangle\otimes|Bell\rangle$ does $B^{(2)}(1,
2;|Bell\rangle\otimes|Bell\rangle)=B^{(2)}(3,
4;|Bell\rangle\otimes|Bell\rangle)=1$ and the others are zero.
$B^{(3)}(1,2,3;|GHZ_3\rangle\otimes |0\rangle)=1$ and otherwise
$0$. Our measure gives the ordering for entanglement degrees in
multiqubit systems, which depends on $m$.

So far we have focused on the entanglement measure of pure
multiqubits. Now we intend to consider whether our measure extends
to higher dimensional Hilbert spaces. The quantum correlation for
eq. (\ref{abc}) is well defined in higher dimensional Hilbert
space if the Pauli's matrices are substituted by the identity
operator and $N^2 -1$ generators of $SU(N)$. Our definition of $M$
applies to the density operator in the mixed state as
$\langle\bullet\rangle=tr(\rho \bullet)$. In the two-qubit mixed
state, we get $B^{(2)}$ for Werner's state;
\begin{equation}
B^{(2)}(1,2;\rho_W)=\frac{1}{9}(4F-1)^2 \label{wern}
\end{equation}
where $F$ is the fidelity for the singlet state. However, we know
that the state in the region $F\leq\frac{1}{2}$, is separable.
$B^{(2)}(1,2;\rho_W)$ must be zero if $F\leq\frac{1}{2}$. $B$ is
not ready to measure the entanglement for mixed states directly.
The measure for the mixed state can be defined by convex roof as
\begin{equation}
B^{(m)}(\alpha,\beta,\cdots\gamma;\rho)=min \{\sum p_i B^{(m)}(\alpha,\beta,\cdots\gamma;|\Psi_i\rangle)
|\sum_i p_i |\Psi_i\rangle\langle\Psi_i|=\rho\}.
\end{equation}

We here present the general measure of entanglement degree for any
quantum states utilizing the expectation values of Pauli matrices
based on the quantum correlations. Our measure gives full
description on the entanglement structure to the given composite
quantum systems. For instance, $B^{(4)}$ has the same magnitude in
both $|GHZ_3\rangle\otimes|0\rangle$ and
$|Bell\rangle\otimes|Bell\rangle$, but $B^{(2)}$ and $B^{(3)}$
present the difference in the entanglement measure of subsystems.
Entanglement ordering for comparing state are different depending
on $m$. We believe that this feature leads us to find mathematical
and physical avenues to construct the classification of multiqudit
entanglements.

\vspace{2.0cm}

\centerline{\bf Acknowledgements}

Lee and Ahn were supported by the Korean Ministry of Science and
Technology through the Creative Research Initiatives Program under
Contact No. M10116000008-02F0000-00610. Oh was supported by Kosef
R06-2002-007-01003-0(2002) and KRF-2002-070-C0029.

\end{document}